\documentstyle[sprocl,diagrams]{article}

\def\be{\begin{equation}}
\def\ee{\end{equation}}
\def\bea{\begin{eqnarray}}
\def\eea{\end{eqnarray}}

\begin{document}

\title{Invariants of multiple-qubit systems under stochastic local operations\\}

\author{G.~S.~JAEGER}

\address {Department of Electrical and Computer Engineering \\
Boston University, Boston 02215\\E-mail: jaeger@bu.edu}

\author{M.~TEODORESCU-FRUMOSU}

\address {Department of Mathematics \\
Boston University, Boston 02215\\E-mail: matf@math.bu.edu}

\author{A.~V.~SERGIENKO}

\address {Department of Electrical and Computer Engineering \\
Boston University, Boston 02215}
\address {Department of Physics\\
Boston University, Boston 02215\\E-mail: alexserg@bu.edu}

\author{B.~E.~A.~SALEH}

\address {Department of Electrical and Computer Engineering \\
Boston University, Boston 02215\\E-mail: besaleh@bu.edu}

\author{M.~C.~TEICH}

\address {Department of Electrical and Computer Engineering \\
Boston University, Boston 02215}
\address {Department of Physics\\
Boston University, Boston 02215\\teich@bu.edu}

\medskip

\maketitle\abstracts{We investigate the behavior of quantum states
under stochastic local quantum operations and classical
communication (SLOCC) for fixed numbers of qubits. We explicitly
exhibit the homomorphism between complex and real groups for
two-qubits, and use the latter to describe the effect of SLOCC
operations on two-qubit states. We find an expression for the
polarization Lorentz group invariant length, which is the
Minkowskian analog of the quantum state purity, the corresponding
Euclidean length. The construction presented is immediately
generalizable to any finite number of qubits.}

\section{Introduction.}

In quantum information theory, stochastic local operations and
classical communication (SLOCC) on single-qubit density matrices
[2] are described by the group SL(2,C), which is homomorphic to
the proper Lorentz group, O$_o$(1,3). The state of a single
classical spin is known to have an invariant length under
transformations of the proper Lorentz group [1]. Here, we consider
the Lorentz-group invariant length for every possible finite
number of qubits, {\it i.e.} quantum spins, which are capable of
being entangled. This length is seen to be the Minkowskian analog
of the quantum state purity, which is the corresponding Euclidean
length. This length is a new tool for describing the behavior of
states of any finite number of qubits under SLOCC, which have thus
far been studied in detail for only two qubits using matrix
methods, which are not obviously generalizable to more than two
qubits but have produced encouraging results [3]. The tensorial
method and results presented here are generalizable to any fixed
number of qubits without difficulty.

\section{A single qubit.}

In classical physics, one can use the expectation values of the
Pauli spin matrices to fully characterize a state of spin, and to
visualize it geometrically via a Poincar\'e sphere. As Han {\it et
al.} [4] have pointed out, these classical parameters form a
Minkowskian four-vector under the group of transformations
corresponding to ordinary and hyperbolic state rotations. In
particular, the elements of the group of proper Lorentz
transformations $O_o(1,3)$ acting on the classical Stokes vector
can be represented as products of the following six forms of
matrix, $M_1, ..., M_6$:

$$M_1(\alpha)=\pmatrix{
  1 & 0 & 0 & 0 \endline
  0 & cos\alpha & -sin\alpha & 0 \endline
  0 & sin\alpha & cos\alpha & 0 \endline
  0 & 0 & 0 & 1}\eqno(1a)$$

\smallskip

$$M_2(\beta)=\pmatrix{
  1 & 0 & 0 & 0 \endline
  0 & cos\beta & 0 & -sin\beta \endline
  0 & 0 & 1 & 0 \endline
  0 & sin\beta & 0 & cos\beta}\eqno(1b)$$

$$M_3(\gamma)=\pmatrix{
  1 & 0 & 0 & 0 \endline
  0 & 1 & 0 & 0 \endline
  0 & 0 & cos\gamma & -sin\gamma \endline
  0 & 0 & sin\gamma & cos\gamma}\eqno(1c)$$

$$M_4(\chi)=\pmatrix{
   cosh\chi & sinh\chi & 0 & 0\endline
   sinh\chi & cosh\chi & 0 & 0\endline
   0 & 0 & 1 & 0\endline
   0 & 0 & 0 & 1}\eqno(1d)$$

$$M_5(\omega)=\pmatrix{
   cosh\omega & 0 & sinh\omega & 0\endline
   0 & 1 & 0 & 0\endline
  sinh\omega & 0 & cosh\omega & 0 \endline
  0 & 0 & 0 & 1}\eqno(1e)$$

$$M_6(\zeta)=\pmatrix{
   cosh\zeta & 0 & 0 & sinh\zeta\endline
   0 & 1 & 0 & 0 \endline
   0 & 0 & 1 & 0 \endline
   sinh\zeta & 0 & 0 & cosh\zeta}\ ,\eqno(1f)$$

\bigskip

\noindent that preserve an associated invariant length ({\it cf.}
[5]). For the investigation of the properties of qubit states, it
is illustrative first to consider Lorentz group transformations in
correspondence to transformations on elements of $\textsl{H}(2)$,
the vector space of all 2x2 complex Hermitian matrices that
includes the density matrices describing states of single qubits.

The state of a quantum ensemble of independent qubits can be
completely described by the set of expectation values
$$x_\mu=Tr(\rho\sigma_\mu)\ \ \ \ (\mu = 0, 1, 2, 3)\ ,\eqno(2)$$

\noindent  where $\sigma_0=\textsf{1}_{2\times 2}$ and $\sigma_i,\
i = 1, 2, 3$, are the Pauli matrices. Likewise, one can write the
density matrix as
$$\rho={1\over 2}\sum_{\mu=0}^3 x_\mu \sigma_\mu, \eqno(3)$$

\noindent and the vector space for one qubit state-vectors is
${\mathcal C}^2.$ Since $\sigma_\mu^2=1$ and ${1\over
2}\sigma_\mu\sigma_\nu =\delta_{\mu\nu}$, the four Pauli matrices
form a basis for \textsl{H}(2) of which the density matrices,
$\rho,$ are the positive-definite, elements of unit trace ({\it
i.e.}, those for which $x_0\equiv 1$), that capture the general
qubit state, pure or mixed.

Now consider these expectation-value vectors in the Minkowskian
real vector space, ${R}^4_{1,3}$, the four-dimensional real vector
space ${ R}^4$ endowed with the Minkowski metric $(+, -, -, -)$,
{\it i.e.} together with a metric tensor $g^{\mu \nu}$ possessing,
as non-zero elements, the diagonal entries $+1, -1, -1$, and $-1$
. The length of a four-vector $x_\mu$ in ${R}^4_{1,3}$ is given by
$<x,x>=g^{\mu\nu}x_\mu x_\nu$. More explicitly, in $R^4_{1,3}$,
the length of a vector $x=(x_0, x_1, x_2, x_3)$ is given by
$$\parallel x\parallel^2_{R^4_{1,3}}=
x_0^2-x_1^2-x_2^2-x_3^2\ \ .\eqno(4)$$

\noindent Using the standard vector basis for ${ R} ^4$,
$e_0=(1,0,0,0)$, $e_1=(0,1,0,0)$, $e_2=(0,0,1,0)$,
$e_3=(0,0,0,1),$  there exists a natural vector-space isomorphism,
$\nu : { R}^4_{1,3}\rightarrow \textsl{H}(2),$ relating the space,
${ R}^4_{1,3}$, of these vectors and the space of state matrices,
\textsl{H}(2), defined by
$$\nu(x_0 , x_1, x_2 , x_3 )=
x_0\sigma_0 + x_1\sigma_1 + x_2\sigma_2 + x_3\sigma_3\ .\eqno(5)$$

\noindent This isomorphism straightforwardly relates the
corresponding basis elements for the space of expectation-value
vectors to those for the space of density matrices, namely
$\nu(e_i)=\sigma_i$ [5]. If we then define the norm on the space
of density matrices, \textsl{H}(2) to be
$$\parallel X\parallel_{H(2)}^2=det X\ ,\ \forall X\in
\textsl{H}(2)\ ,\eqno(6)$$

\noindent then the isomorphism $\nu$ between the spaces of these
real vectors and the Hermitian matrices becomes  a
length-preserving mapping, {\it i.e.} an isometry, since we have
the following simple relationship between lengths in the two
spaces:
$$\parallel \nu(x_0, x_1, x_2, x_3)\parallel_{H(2)}^2\equiv det X=
x_0^2-x_1^2-x_2^2-x_3^2=\parallel x\parallel^2_{{ R}_{1,3}^4}\
.\eqno(7)$$

\noindent  Since the Pauli matrices are traceless and
$\sigma_\mu^2 =1_2\ \ (\mu = 0, 1, 2, 3)$, we obtain the following
expression for the inverse, $\nu^{-1}: \textsl{H}(2)\rightarrow {
R}^4_{1,3}$, of this vector-space isomorphism:
$$\nu^{-1}(X)={1\over 2}\bigg(Tr(X), Tr(X\sigma_1), Tr(X\sigma_2),
Tr(X\sigma_3)\bigg)\ ,\ \forall X\in \textsl{H}(2) ,\eqno(8)$$

\noindent which maps the space of $2\times 2$ Hermitian matrices
containing the density matrices into the space, ${ R}_{1,3}^4$,
containing the quantum four-vectors. In particular, the density
matrices of quantum mechanics are identified within the space of
Hermitian matrices \textsl{H}(2) as those having trace one, a
condition guaranteeing that the sum of probabilities of all the
possible events for the quantum state is unity.

Defining the contraction map $\lambda: \textsl{H}(2)\rightarrow
\textsl{H}(2)$:
$$\lambda(X)={1\over 2}X\ , \forall X\in \textsl{H}(2)\ ,$$

\noindent allows us to define the isomorphism
$\omega(x)=\lambda\circ\nu: { R}^4_{1,3}\rightarrow \textsl{H}(2)$
of the space containing expectation-value vectors to that
containing the density matrices:
$$\omega(x_0, x_1, x_2, x_3)={1\over 2}\sum_{i=0}^3 x_\mu\sigma_\mu\
\ ,\ \forall x=(x_0, x_1, x_2, x_3)\in R^4_{1,3}\ .\eqno(9)$$

\noindent The corresponding inverse map,
$\omega^{-1}:\textsl{H}(2)\rightarrow R^4_{1,3}$ is
$$\omega^{-1}(X)=\bigg(Tr(X), Tr(X\sigma_1), Tr(X\sigma_2),
Tr(X\sigma_3)\bigg)\ .\eqno(10)$$

\noindent As with $\nu$, $\omega$ becomes an isometry if we define
$\parallel\omega(x_0, x_1, x_2, x_3)\parallel^2_{H(2)}\equiv
det(2X)=\parallel x\parallel^2_{ R^4_{1,3}}\ .$

\smallskip
\noindent $\omega^{-1}$ now directly returns the vector of
expectation values, $x_\mu =Tr (\rho\sigma_\mu)$ ($\mu = 0, 1, 2,
3$), as desired.

The group action $\alpha$:
SL(2,C)$\times$\textsl{H}(2)$\rightarrow$ \textsl{H}(2), on $H(2)$
is defined by
$$\alpha(A,X)=AXA^*,\ \ \forall A\in
SL(2,C)\ \ and\ \ \forall X\in \textsl{H}(2)\ ,\eqno(11)$$

\noindent involving the density matrices. We see that the norm
induced by the isomorphism $\omega$ is preserved under $\alpha$,
since
$$\parallel AXA^*\parallel_{H(2)}^2= det(AXA^*)
=|detA|^2detX=detX=\parallel X\parallel_{H(2)}^2.\eqno(12)$$

\noindent The natural group action, ${\beta}:O_o(1,3)\times{
R}_{1,3}^4\rightarrow { R}_{1,3}^4$, of the Lorentz group
$O_0(1,3)$ on the quantum observables, the elements of ${
R}_{1,3}^4$ including the vectors describing this ensemble is
defined by
$$\beta(x)=Bx ,\ \forall B\in O_o(1,3),\ \ \forall x\in { R}_{1,3}^4 \ ,\eqno(13)$$

\noindent and is norm-preserving (by definition), {\it i.e.}
$\parallel Bx\parallel^2_{R_{1,3}^4}=\parallel x\parallel^2_{
R_{1,3}^4}$.

Since the isomorphism $\omega$ of the expectation value space to
the space containing the quantum states is an isometry, we can
also define a map, ${\theta}:SL(2,C)\rightarrow O_0(1,3)$, between
the transformations on elements of \textsl{H}(2), including the
density matrices, to those transformations of elements of ${
R}_{1,3}^4$. The action of a matrix $A$ on the matrices $X\in
\textsl{H}$(2) induces a corresponding Lorentz transformation
$\theta(A)$ of vectors in $R_{1,3}^4$, such that $\parallel
\omega^{-1}(AXA^*)\parallel_{R_{1,3}^4}=
\parallel\theta(A)\omega^{-1}(X)\parallel_{R_{1,3}^4}=
\parallel\omega^{-1}(X)\parallel_{R_{1,3}^4}\ .\\ $

\smallskip

By defining a map, $\gamma$, of the quantum state transformations
into the corresponding transformations of the qubit expectation
values, $\gamma : SL(2,C)\times H(2)\rightarrow O_o(1,3)\times{
R}_{1,3}^4$,
$$\gamma(A,X)=\bigg(\theta(A),\omega^{-1}(X)\bigg)\ , \eqno(14)$$

\noindent we then obtain a commuting diagram, {\it i.e.} a set of
mathematical objects and mappings such that any two mappings
between any pair of objects obtained by composition of mappings
are equal. This illustrates in full detail the well-known
relationship between SL(2,C) and O$_0$(1,3), but tailored to the
quantum mechanical context.

\def\Al{\alpha}\def\Be{\beta}\def\Ga{\gamma}\def\Om{\omega}
\begin{diagram}
SL(2,C)\times\textsl{H}(2)
&\rTo^\Al & \textsl{H}(2)\\
\dTo^\Ga &  & \uTo^\Om\\
O_o(1,3)\times{ R}_{1,3}^4 & \rTo^\Be & { R}_{1,3}^4\\
\end{diagram}

\bigskip

\noindent  The above construction allows one to freely analyze the
behavior of the quantum expectation values under Lorentz group
transformations and will be generalized below.

The Minkowskian length, $l^2$, of the vector of expectation values
is
$$l^2=x_0^2-x_1^2-x_2^2-x_3^2,\eqno(15)$$

\noindent following Eq. (4), being similar to its analog in the
classical realm and invariant under the Lorentz group of
transformations represented by the basic forms $M_1, ..., M_6$.
This group of transformations goes beyond the limited context of
unitary transformations of density matrices (for which $x_0\equiv
1$), to include non-unitary transformations (for example,
corresponding to the Lorentz group transformations $M_4, M_5,
M_6$). The loci of constant $l^2$ are three-dimensional
hyperboloids -- a range of ensemble relative sizes $x_0$ and
polarization vector states - lying within what is the probability
analog of the ``forward light cone" of special relativity.

When the corresponding transformation of the density matrix is an
element of the SU(2) subgroup of SL(2,C), corresponding to a
unitary transformation of density matrices into density matrices
and $x_0$ is strictly unity, the states lie within a locus a fixed
distance from the $x_0$ axis; when this transformation involves
one of $M_4, M_5, M_6$ probability is lost/gained, so that this
constraint is no longer obeyed and $x_0$ can take other values,
$x_0'$, and move to other locations within the hyperboloid
represented by the same value of the invariant.

\section{More than one qubit.}

To show how we can apply in a well-defined way Lorentz
transformations to multiple qubit systems, including those that
are entangled, consider now the application of the Lorentz group
to two-qubit systems. We introduce the joint expectation values
$x_{\mu\nu}=Tr(\rho\sigma_\mu\otimes\sigma_\nu)$, where $\mu ,
\nu= 0, 1, 2, 3$, and express the matrix of the general state of a
two qubit ensemble [6,7]:
$$\rho={1\over 4}\sum_{\mu,\nu=0}^3x_{\mu\nu}\sigma_\mu\otimes\sigma_\nu\
,\eqno(16)$$

\noindent where $\sigma_\mu\otimes\sigma_\nu$ ($\mu , \nu = 0, 1,
2, 3$) are simply tensor products of the identity and Pauli
matrices, and the state-vector space for pure states of two qubits
is ${\mathcal C}^2\otimes{\mathcal C}^2$. The four-vector,
$x_\mu$, must then be generalized to a 16-element tensor,
$x_{\mu\nu}$.

The two-qubit density matrices $\rho$ are positive, unit-trace
elements of the 16-dimensional complex vector space of Hermitian
$4\times 4$ matrices, \textsl{H}(4). The tensors
$\sigma_\mu\otimes\sigma_\nu\equiv\sigma_{\mu\nu}$ provide a basis
for \textsl{H}(4), which is isomorphic to the tensor product space
\textsl{H}(2)$\otimes\textsl{H}$(2) of the same dimension, since
${1\over
4}Tr(\sigma_{\mu\nu}\sigma_{\alpha\beta})=\delta_{\mu\alpha}\delta_{\nu\beta}$
and $\sigma_{\mu\nu}^2=1_{2\times 2}$, in analogy to the
single-qubit case. We can write the two-qubit expectation values
as
$$x_{\mu\nu}=Tr(\rho\ \sigma_\mu\otimes\sigma_\nu).\eqno(17)$$

\bigskip

\noindent A density matrix for the general state of a two-qubit
system is thus an element of $\textsl{H}(4)\simeq H(2)\otimes
\textsl{H}(2)$ of the form
$$\rho={1\over
4}\bigg(\sigma_0\otimes\sigma_0+\sum_{i= 1}^3
x_{i0}\sigma_i\otimes\sigma_0 + \sum_{j=
1}^3x_{0j}\sigma_0\otimes\sigma_j +\sum_{i, j=
1}^3x_{ij}\sigma_i\otimes\sigma_j\bigg)\ ,\eqno(18)$$

\noindent an element of the Hilbert-Schmidt space [7] that
corresponds to
$$x=\ e_0\otimes e_0
+\sum_{i= 1}^3x_{i0}\ e_i\otimes e_0 + \sum_{j= 1}^3x_{0j}
e_0\otimes e_j + \sum_{i, j= 1}^3x_{ij}\ e_i\otimes e_j\
\eqno(19)$$

\noindent in $R_{1,3}^4\otimes{ R}_{1,3}^4,$ expressed in terms of
the elements of standard vector basis for ${ R} ^4$,
$e_0=(1,0,0,0)$, $e_1=(0,1,0,0)$, $e_2=(0,0,1,0)$,
$e_3=(0,0,0,1).$

The isomorphism between the space of two-qubit expectation values
and two-qubit density matrices, $\omega\otimes\omega$:
$R_{1,3}^4\otimes R_{1,3}^4\rightarrow H(2)\otimes H(2)\simeq
H(4)$ is defined as
$$(\omega\otimes\omega)(v\otimes w)\equiv\omega(v)\otimes\omega(w)\ ,\eqno(20)$$

\noindent for all $v,w\in{ R}_{1,3}^4.$ $\sigma_\mu\otimes
\sigma_\nu$ form a basis for the required space of two-qubit
Hermitian matrices $H(2)\otimes H(2)\simeq H(4)$, and
$(\omega\otimes\omega)(e_\mu\otimes e_\nu)=
\omega(e_\mu)\otimes\omega(e_\nu)=\sigma_\mu\otimes\sigma_\nu$
($\mu , \nu = 0, 1, 2, 3$). Furthermore, the inverse map taking
density matrices to two-qubit tensors,
${(\omega\otimes\omega)}^{-1}:\textsl{H}(4)\rightarrow R_{1,3}^4
\times R_{1,3}^4$ is given by
$${(\omega\otimes\omega)}^{-1}(X)=Tr(X\sigma_\mu\otimes\sigma_\nu )
\ ,\ \eqno(21)$$ \noindent for all $X\in H(4).$ To describe the
effect of the full set of group transformations, we use the map
$\alpha\otimes\alpha : SL(2,C)\times SL(2,C)\times\textsl{H}(2)
\otimes \textsl{H}(2) \rightarrow \textsl{H}(2)\otimes
\textsl{H}(2),$ since for each qubit the group of transformations
$SL(2,C)$ acts via the action $\alpha$ on the vector space
\textsl{H}(2) that includes the density matrices. The action on
the two-qubit Hermitian matrices is defined as
$$(\alpha\otimes\alpha)(A,B,X\otimes
Y)=(AXA^*)\otimes(BYB^*),\eqno(22)$$

\noindent for all $(A,B)\in SL(2,C)\times SL(2,C),$ and $\forall
X, Y\in$ \textsl{H}(2).\ The action $\alpha\otimes\alpha$ is
norm-preserving on the tensor-product space, since
$$\parallel AXA^*\parallel^2_{H(2)}=\parallel
X\parallel^2_{H(2)}\eqno(23a)$$

$$\parallel BYB^*\parallel^2_{H(2)}=\parallel
Y\parallel^2_{H(2)}\ .\eqno(23b)$$

The action $\beta$ of the Lorentz group $O_o(1,3)$ on the space of
expectation values, ${ R}_{1,3}^4$, also generalizes in the
two-qubit case to $\beta\otimes\beta: O_o(1,3)\times
O_o(1,3)\times{ R}_{1,3}^4 \otimes{ R}_{1,3}^4 \rightarrow{
R}_{1,3}^4\otimes{ R}_{1,3}^4,$

$$(\beta\otimes\beta)((C,D), v\otimes w)=(Cv)\otimes(Dw)\
\eqno(24)$$

\bigskip

\noindent for all $(C,D)\in O_o(1,3)\times O_o(1,3)$ and $\forall
v\otimes w\in R_{1,3}^4 \otimes R_{1,3}^4.$

The isomorphism $\omega\otimes\omega$ is an isometry, so we define
the group homomorphism $\theta\times\theta: SL(2,C)\times
SL(2,C)\rightarrow O_o(1,3)\times O_o(1,3)$. The action of the
transformations $A\times B\in SL(2,C)\times SL(2,C)$ on the
matrices $X\otimes Y\in\textsl{H}(2)\otimes\textsl H(2)$, which
include the density matrices, induces a corresponding Lorentz
group transformation $\theta(A)\times\theta(B)$ on the space of
expectation-value tensors $R_{1,3}^4 \otimes R_{1,3}^4$:

$$(\omega\otimes\omega)^{-1}
[(AXA^*)\otimes(BYB^*)]=\omega^{-1}(AXA^*)\otimes\omega^{-1}(BYB^*)$$

$$\ \ \ \ \ \ \ \ \ \ \ \ \ \ \ \ \ \ \ \ \ \ \ \ \ \  \ \ \ \ \ \ \ \ \ \ \ \
=\theta(A)\omega^{-1}(X)\otimes\theta(B)\omega^{-1}(Y)\
.\eqno(25)$$

\noindent The $\theta(A)\times\theta(B)$ are well-defined Lorentz
group transformations since, as before,

\smallskip

$\ \ \ \ \ \ \parallel\omega^{-1}(AXA^*)\parallel^2_{R_{1,3}^4}
=\parallel\theta(A)\omega^{-1}(X)\parallel^2_{R_{1,3}^4}
=\parallel\omega^{-1}(X)\parallel^2_{R_{1,3}^4}$

\noindent and

$\ \ \ \ \ \ \parallel\omega^{-1}(BYB^*)\parallel^2_{R_{1,3}^4}
=\parallel\theta(B)\omega^{-1}(Y)\parallel^2_{R_{1,3}^4}
=\parallel\omega^{-1}(Y)\parallel^2_{R_{1,3}^4}\ .$

Defining the map acting on the space $H(2)\otimes H(2)$ including
the density matrices,

$\gamma\otimes\gamma:SL(2,C)\times SL(2,C)\times H(2)\otimes
H(2)\rightarrow O_o(1,3)\times O_o(1,3)\times R_{1,3}^4\otimes{
R}_{1,3}^4$,

\noindent by

$$(\gamma\otimes\gamma)\bigg((A,B),(X\otimes Y)\bigg)=
\Bigg(\bigg((\theta\times\theta)(A,B)\bigg),
(\omega\otimes\omega)^{-1}(X\otimes Y)\Bigg)$$

$$\ \ \ \ \ \ \ \ \ \ \ \ \ \ \ \ \ \ \ \ \ \ \ \ \ \ \ \ \
=\Bigg(\bigg(\theta(A),\theta(B)\bigg),
\omega^{-1}(X)\otimes\omega^{-1}(Y)\Bigg),\eqno(26)$$

\noindent for all $(A,B)\in SL(2,C)\times SL(2,C)$ and for all
$X\otimes Y\in\textsl{H}(2)\otimes\textsl{H}(2)$, we obtain the
following commuting diagram

\bigskip

\def\Al{\alpha\otimes\alpha}\def\Be{\beta\otimes\beta}\def\Ga{\gamma\otimes\gamma}
\def\Om{\omega\otimes\omega}
\begin{diagram}
SL(2,C)\times SL(2,C)\times\textsl{H}(2)\otimes\textsl{H}(2)
&  \rTo^\Al & \textsl{H}(2)\otimes\textsl{H}(2)\\
\dTo^\Ga & & \uTo^\Om\\
O_o(1,3)\times O_o(1,3)\times R_{1,3}^4\otimes R_{1,3}^4 &
\rTo^\Be & R_{1,3}^4\otimes R_{1,3}^4\\
\end{diagram}

\bigskip

\noindent demonstrating the well-definedness of the construction
on a set of two-qubit states, including those that are entangled.

Again the length given by the tensor norm
$$l_{12}^2\equiv\parallel x\parallel^2_{R_{1,3}^4\otimes
R_{1,3}^4}=<x,x>=(x_{00})^2-\sum_{i=1}^3(x_{i0})^2-\sum_{j=1}^3(x_{0j})^2
+\sum_{i=1}^3\sum_{j=1}^3(x_{ij})^2\ .\ \ \ \ \ \ \ \ \eqno(20)$$

\smallskip

\noindent is invariant under Lorentz group transformations
$(A,B)\in O_o(1,3)\times O_o(1,3)$. A similar approach can be used
to find an expression for this length for an arbitrary number of
qubits.

The extension of the above approach to the case of n-qubits is
straightforward, and allows us to find the invariant length for
any finite number of qubits. Unlike previous approaches to
applying the Lorentz group to quantum states (such as that of Ref.
[3]) that used matrix methods to arrive at quantities of interest,
the approach of the present treatment is manifestly general.

The n-qubit tensor $x_{i_1 ...i_n}$ transforms under the group
O$_o(1,3)$ as

$$x'_{i_1... i_n}=\sum_{j_1,... , j_n =0}^3L^{j_1}_{i_1}... L^{j_n}_{i_n}x_{j_1...j_n}\ ,\eqno(37)$$

\noindent where the $L^{j}_{i}$ are such transformations acting in
the spaces of qubits $1, ..., n$. Again, each such transformation
$x_{\mu_1\mu_2... \mu_n}\rightarrow x_{\mu_1\mu_2... \mu_n}'$ of a
given n-qubit expectation-value tensor will yield a new Hermitian
state matrix $\rho'$. After transformation, the tensor element
$x_{0...0}'$ is the new n-qubit ensemble relative size. Again, the
renormalizing of $\rho'$ gives the resulting density matrix for
the ensemble: $\rho''=\rho'/Tr(\rho')$.

Note that the quantum state purity $Tr\rho^2$ for a general
n-photon state,

$$Tr\rho^2 =Tr\bigg[\bigg({1\over 2}\bigg)^n\sum_{i_1,
...,i_n=0}^3 x_{i_1 ... i_n}\sigma_{i_1}\otimes
...\otimes\sigma_{i_n}\ \ \bigg({1\over 2}\bigg)^n\sum_{j_1,
...,j_n=0}^3 x_{j_1 ... j_n}\sigma_{j_1}\otimes
...\otimes\sigma_{j_n}\bigg]\ ,\eqno(38)$$

\medskip

\noindent has a particularly simple form in terms of the elements
n-qubit four-tensor; since
$(\sigma_{i_1}\otimes\sigma_{j_1})\otimes
...\otimes(\sigma_{i_n}\otimes\sigma_{j_n})=\sigma_0\otimes
...\otimes\sigma_0$ if and only if $i_k=j_k$, for all $k=1, 2,
..., n$,  only the coefficient of the term $\sigma_0\otimes
...\otimes\sigma_0$ contribute to the trace, and we have

$$Tr\rho^2={1\over 2^n}\sum_{i_1, ..., i_n=0}^3x^2_{i_1 ... i_n}\ .\eqno(39)$$

\noindent The state purity is thus seen to be the Euclidean analog
of the Minkowskian invariant length.

\medskip

\section{Conclusion.}

We have considered the application of the Lorentz group to
multiple-qubit states. We have exhibited the necessary
construction for two-qubit case in detail. We showed that the
multiple qubit state expectation values form Minkowskian tensors
with a related invariant length under the action of the Lorentz
group. This length is the Minkowskian analog of the quantum state
purity, which is the corresponding Euclidean length. This length
provides a new tool for describing the behavior of states of any
finite number of qubits under SLOCC, including those in entangled
states, which have thus far been studied with positive results but
for only two-qubit states and two-qubit reduced states of
three-qubit pure states [3]. We conjecture that the SLOCC
invariant length describes entanglement properties of multiple
qubit states.

\end{document}